\documentclass{article}
\input amssym.def
\input amssym.tex
\def\ben{\begin{equation}}
\def\een{\end{equation}}
\def\half{{1 \over 2}}
\def\bea{\begin{eqnarray}}
\def\eea{\end{eqnarray}}

\def\ft#1#2{{\textstyle{{\scriptstyle #1}\over {\scriptstyle #2}}}}
\def\fft#1#2{{#1 \over #2}}

\def\half{{\textstyle{1\over2}}}

  \let\n=\nu   
     
       \let\D=\Delta  
   \let\U=\Upsilon  
\let\C=\Chi

 \def\bd{\begin{document}} \def\ed{\end{document}}
\def\ds{\documentstyle} \let\fr=\frac \let\bl=\bigl \let\br=\bigr
\let\Br=\Bigr \let\Bl=\Bigl 
\let\bm=\bibitem
\let\na=\nabla
\let\pa=\partial 
\let\ov=\overline 
\newcommand{\be}{\begin{equation}} 
\newcommand{\ee}{\end{equation}} 
\def\ba{\begin{array}}
\def\ea{\end{array}}
\def\ft#1#2{{\textstyle{{\scriptstyle #1}\over {\scriptstyle #2}}}}
\def\fft#1#2{{#1 \over #2}}
\def\del{\partial}
\def\vp{\varphi}
\def\sst#1{{\scriptscriptstyle #1}}
\def\oneone{\rlap 1\mkern4mu{\rm l}}
\def\td{\tilde}
\def\wtd{\widetilde}
\def\ie{\rm i.e.\ }
\def\dalemb#1#2{{\vbox{\hrule height .#2pt
        \hbox{\vrule width.#2pt height#1pt \kern#1pt
                \vrule width.#2pt}
        \hrule height.#2pt}}}
\def\square{\mathord{\dalemb{6.8}{7}\hbox{\hskip1pt}}}
\newcommand{\ho}[1]{$\, ^{#1}$}
\newcommand{\hoch}[1]{$\, ^{#1}$}  
\newcommand{\ra}{\rightarrow}
\newcommand{\lra}{\longrightarrow}
\newcommand{\Lra}{\Leftrightarrow}
\newcommand{\ap}{\alpha^\prime}
\newcommand{\bp}{\tilde \beta^\prime}
\newcommand{\tr}{{\rm tr} }
\newcommand{\Tr}{{\rm Tr} } 
\def\0{{\sst{(0)}}}
\def\1{{\sst{(1)}}}
\def\2{{\sst{(2)}}}
\def\3{{\sst{(3)}}}
\def\4{{\sst{(4)}}}
\def\5{{\sst{(5)}}}
\def\6{{\sst{(6)}}}
\def\7{{\sst{(7)}}}
\def\8{{\sst{(8)}}}
\def\n{{\sst{(n)}}}
\def\cA{{{\cal A}}}
\def\cF{{{\cal F}}}
\def\tV{\widetilde V}
\def\tW{\widetilde W}
\def\tH{\widetilde H}
\def\tE{\widetilde E}
\def\tF{\widetilde F}
\def\tA{\widetilde A}
\def\im{{{\rm i}}}
\def\jm{{{\rm j}}}
\def\km{{{\rm k}}}
\def\tY{{{\wtd Y}}}
\def\ep{{\epsilon}}
\def\vep{{\varepsilon}}
\def\R{\rlap{\rm I}\mkern3mu{\rm R}}
\def\bD{{{\bar D}}}
\def\C{{{\Bbb C}}}
\def\H{{{\Bbb H}}}
\def\RP{{{\Bbb R}{\Bbb P}}} 
\def\CP{{{\Bbb C}{\Bbb P}}} 
\def\HP{{{\Bbb H}{\Bbb P}}}
\def\Z{{\Bbb Z}} 
\def\bfe{{\bf e}}
\def\bfq{{\bf q}}
\def\cosec{\rm cosec}

\begin{document}

\newcount\hour \newcount\minute
\hour=\time  \divide \hour by 60
\minute=\time
\loop \ifnum \minute > 59 \advance \minute by -60 \repeat
\def\nowtwelve{\ifnum \hour<13 \number\hour:
                      \ifnum \minute<10 0\fi
                      \number\minute
                      \ifnum \hour<12 \ A.M.\else \ P.M.\fi
	 \else \advance \hour by -12 \number\hour:
                      \ifnum \minute<10 0\fi
                      \number\minute \ P.M.\fi}
\def\nowtwentyfour{\ifnum \hour<10 0\fi
		\number\hour:
         	\ifnum \minute<10 0\fi
         	\number\minute}
\def\now{\nowtwelve}

\newcommand{\labeq}[1] {\label{eq:#1}}
\newcommand{\eqn}[1] {(\ref{eq:#1})}
\newcommand{\labfig}[1] {\label{fig:#1}}
\newcommand{\fig}[1] {\ref{fig:#1}}

\title{Birkhoff's invariant and Thorne's Hoop Conjecture}
\author{G. W. Gibbons 
\\D.A.M.T.P.,
\\Cambridge,
\\Wilberforce Road,
\\Cambridge CB3 0WA,
\\U.K.}

\maketitle
\medskip
\medskip
\begin{abstract}
I propose a sharp form of Thorne's hoop conjecture
which relates Birkhoff's invariant $\beta$ for  an outermost
apparent horizon  to its $ADM$ mass,  $ \beta \le 4 \pi M_{ADM}$.
I prove the conjecture in the case of collapsing null shells
and provide further evidence from exact 
rotating black hole solutions.  Since $\beta$ is bounded
below by the length $l$ of the  shortest 
non-trivial geodesic lying in the apparent horizon, the conjecture
implies $l \le 4 \pi M_{ADM}$. 
The Penrose conjecture, $\sqrt{\pi A} \le 4 \pi M_{ADM}$, and Pu's theorem 
imply this latter consequence for horizons 
admitting an antipodal isometry. Quite generally, Penrose's inequality
and Berger's   isembolic inequality,  $\sqrt{\pi A}
 \ge { 2 \over \sqrt \pi} i$,
where $i$ is the injectivity radius, imply
$ 4c  \le 2 i  \le 4 \pi M_{ADM}$, where $c$ is the convexity radius.

\end{abstract}
\vfill 
\vfill 

\section{The Hoop Conjecture}
Thorne's  original  Hoop Conjecture \cite{Thorne}   was that

\begin{quote}{ \narrower Horizons form when and only when a mass $m$ gets
compacted into a region whose circumference in EVERY direction is
$C\le 4 \pi M$. }\end{quote} 

The capitalization  \lq \lq EVERY \rq \rq was 
intended to emphasis the fact that while the collapse of 
oblate shaped bodies  the circumferences are all roughly equal,
in the  prolate case, a the collapse of a long almost cylindrically 
shaped body whose girth was nevertheless small would not necessarily
produce a horizon.  However,  as  proposed, the
statement is so imprecise as  to render  either proof or disproof
impossible.  Presumably for the  mass we could take the ADM mass, $M_{ADM}$,
but what about the circumference of the hoop?
Since Thorne's article, there have been many attempts
to tighten  up the formulation of the conjecture,
the most recent of which is \cite{Senovilla}
to which the reader is referred for an extensive list of 
previous contributions to this question.   
 
The present note was inspired by \cite{Paiva} 
and  in many respects takes forward a suggestion of Tod \cite{Tod}.
It is closely related to recent work of Yau and others on
the idea of quasi-local mass.

\section{Birkhoff's Invariant and Birkhoff's Theorem}

We can assume that the apparent horizon is topologically 
spherical \cite{Hawking1,Gibbons1,Gibbons2,Hawking2}.
In what follows we follow \cite{Paiva} fairly closely. 
Suppose that $S=\{S^2,g \}$ is a sphere 
with arbitrary metric $g$ and $f:S \rightarrow {\Bbb R}$ a function on
$S$
with just two critical points, a maximum and a minimum.
Each level set $f^{-1}(c),\,c \in {\Bbb R}$ has a length $l(c)$
and for any given function $f$ we define 
\ben
\beta (f) = {\rm max}_c\, l(c) \,.
\een
We now define the Birkhoff invariant $\beta(S,g)$
by minimising $\beta(f)$ over all such  functions  
\ben
\beta = \inf _f \beta (f) \,.
\een

The intuitive meaning of $\beta$ is the least length of a 
length of a closed (elastic) string  or rubber band 
which may be slipped over over the surface
$S$
\cite{Birkhoff}. To understand why, note that each function
$f$ gives a foliation of $S$ by a one parameter family of 
curves $f=c$ which 
we may think of as the string or rubber band at each 
\lq \lq moment of  time \rq\rq\,  $c$.    
$\beta(f)$ is the longest length of the band during that process.
If we change the foliation we can hope to reduce  this longest length
and the infinum is the best that we can do. 
The phrase  \lq \lq moment of  time \rq \rq  is in quotation marks
because we are not regarding $f$ as a physical  time function, 
merely a convenient  way of thinking about the geometry of $S$.

Birkhoff's Theorem \cite{Birkhoff}  then assures us that 
there exist a closed geodesic $\gamma$ on $S$ with  length
$l(\gamma) = \beta (g)$. Clearly, if $l(g)$ is the length of the smallest
non-trivial closed (i.e  periodic) geodesic then
\ben
l(g) \le \beta(g) \,. \label{upper} 
\een

It seems therefore that the Birkhoff invariant $\beta(g)$ should be 
taken as a precise formulation of Thorne's rather  vague notion
of circumference. We shall proceed on this basis. Thus we make the following 

\begin{quote}{\narrower  {\it Conjecture:} For an outermost  marginally
trapped
surface $S$ lying in a Cauchy hypersurface surface $\Sigma$
with ADM mass $M_{\rm ADM}$  on which the Dominant 
Energy condition holds, then
\ben
\beta(g) \le 4 \pi M_{\rm ADM} \,.
\label{conjecture} \een } \end{quote}
 
In other words, (\ref{conjecture}) is conjectured to be a
 necessary condition
for a marginally outermost  trapped surface.
Bearing in mind Thorne's comments about
very prolate shaped surfaces  for which $\beta(g)$ can be extremely
small,  it is not claimed that (\ref{conjecture}) is a sufficient
condition for a closed  surface $S$ to be trapped or marginally.

Clearly, from (\ref{upper}),  this form of the hoop conjecture
implies  
\ben
l(g)  \le 4 \pi M_{\rm ADM} \,,
\label{consequence} \een  
and therefore a counter example to (\ref{consequence}) would be a counter example
to (\ref{conjecture}). 

\subsection{The Kerr-Newman  Horizon}

We first test the conjecture on the general charged
rotating black hole. In standard notation, the metric
on the horizon  is \cite{Smarr}   
\ben
g= ds ^2 = (r_+^2 + a ^2 ) \Bigl 
( ( 1- x^2 \sin ^2 \theta) d \theta ^2  +{
 \sin ^2 \theta d \phi^2 \over 1- x^2 \sin ^2 \theta } 
\Bigr ) \,,   
\een
with
\ben
x^2 = { a^2 \over r_ +^2 + a ^2 } \,. 
\een

This is clearly foliated by the orbits of the
group of rotations generated by by ${\partial \over \partial \phi}$
and we   take $f=\cos \theta$. That is, we are thinking of the coordinate
$\theta$ as a function on $S$. In this case the greatest length of the 
small circles, i.e. of the orbits,  is  $l_e$, 
the length of the  of the equatorial  geodesic at 
$\theta = {\pi \over 2}$ and
we have
\ben
\beta(\cos \theta) = l_e=  2 \pi (  r_+ + { a^2 \over r_+} )  = 
2 \pi ( 2 M - {Q^2 \over r_+} )  \le 4 \pi M  \,.   \label{Kerr-Newman} 
\een
The right hand side of (\ref{Kerr-Newman}) is certainly an upper bound
for the Birkhoff invariant and so the conjecture certainly holds
in this case. 

However the horizon is prolate in character, in the sense that
the polar circumference $l_p$ which is the length of a meridional 
 geodesic $l_p$ (i.e. one with $\phi ={\rm constant}$ and $\phi={\rm
 constant} + \pi$ ), is
\ben
l_p=  \sqrt{r_+^2 + a^2 } \int^\pi _0 \sqrt{1-x^2 \sin^2 \theta }\, d
\theta  
 \,.\een
In fact taking $f=\sin \theta \cos \phi$, we have 
\ben
\beta(\sin \theta \cos \phi) = l_p \,,  
\een
and since
\ben
 \sqrt{r_+^2 + a^2 } \le r_+ + { a^2 \over r_+} \,,
\een
we have 
\ben
\beta(g) \le l_p \le l_e \le 4 \pi M\,.
\een
Despite being prolate, the Gaussian curvature $K$ of the surface is
given by 
\ben
K= { ( {r_+} ^2 + a^2 )  
( r_+ ^2 -3 a ^2 \cos ^2 \theta )  
\over ( { r_+} ^2 + a ^2 \cos ^2 \theta )  ^3 } 
\,,  
\een 
and can become negative at the poles $\theta=0, \pi$.  

The Kerr-Newman metrics have been generalised to include up
to four different charges associated with four different abelian vector
fields \cite{Cvetic}. In the subclass for which
only two charges are non-vanishing we can use the results
of \cite{Chong} to examine the conjecture.
The energy momentum tensor of the system satisfies the Dominant Energy
Condition and the horizon geometry may be extracted from eqn(45)
of \cite{Chong}
\ben
ds ^2 = W d \theta ^2 + { (r_{+1}r_{+2} + a^2 )^2 \over W } 
\sin ^2 \theta  \,d \phi ^2 \,, 
\een     
with
\ben
W= r_{+1}r_{+2} + a^2 \cos^2 \theta \,.
\een
and
\ben
 r_{+1}= r_+ + 2  m \sinh ^2  \delta  _1 \,, 
\qquad  r_{+2}= r_+ + 2 m \sinh ^2  \delta _2
\een
with $r_+$ the larger root of $r^2 -2mr + a^2 =0$
and $\delta _1 $ and $\delta_2$ two parameters specifying the two charges.
If $\delta_1=\delta _2$ we obtain the Kerr-Newman case.

Just as the  horizon geometry of the Kerr-Newman solution
 is isometric to that of the neutral Kerr, so in this more general
case, we find an isometric horizon geometry.
Of course the interpretation of the parameters occurring in the metric
is different,
but the geometry is the same. Thus
\ben
\beta(g) \le l_p \le l_e =  2 \pi \bigl(  \sqrt{r_{+1}r_{+2}}  + { a^2 
\over \sqrt{ r_{+1}r_{+2}}  } \bigr ) \,.  
\een

Now for positive $x,y, z$, 
\ben
xy \le {1 \over 4} (x+y)^2 \,,\quad \Longrightarrow \quad  \sqrt{(z+x)(z+y)}
\le z+ \half (x+y) \,. 
\een
Thus, \ben
\sqrt{ r_{+1}r_{+2}}   \le r_+ + m \bigl
( \sinh ^2  \delta _1 +  \sinh ^2  \delta _2 \bigr) 
\een
and
\ben
{a^2 \over\sqrt { r_{+1}r_{+2}}  }  \le {a^2 \over r_+} \,,   
\een
Thus
\ben
\bigl(  \sqrt{r_{+1}r_{+2}}  + { a^2 
\over \sqrt{ r_{+1}r_{+2}}  } \bigr ) \le  r_+ + m \bigl
( \sinh ^2  \delta _1 +  \sinh ^2  \delta _2 \bigr) + {a^2 \over r_+}.
\een 

But
\ben
2m = r_+ +{a^2 \over r_+} \,,
\een
and the ADM mass is given by
\ben
M_{ADM} = 2m + 2 m \bigl
( \sinh ^2  \delta _1 +  \sinh ^2  \delta _2 \bigr) 
\een
Thus 
\ben
\beta(g) \le 4 \pi M_{ADM} \,,
\een
and the conjecture holds in this case.
It would be interesting to check it in the  four charge case, but the algebra
appears to be rather more complicated.

\section{Collapsing Shells and Convex Bodies}   

There is a class of examples \cite{Gibbons} in which a shell of null matter
collapses at the speed of light in which the apparent horizon
$S$ may be thought of as a convex body isometrically embedded
in  Euclidean space ${\Bbb E} ^3$. In this case one has 
\ben
8 \pi M_{\rm ADM} \ge         \int _S H d A \,, \label{shell}
\een
where $H= \half ({1 \over R_1} + {1 \over R_2 } ) $ is the mean curvature
and $R_1$ and $R_2$ the principal radii of curvature
of $S$  and $dA$ is the area element on $S$.
The right hand side is called the total mean curvature and 
it was  shown by \'Alvarez Paiva    \cite{Paiva}   in this case that
\ben
\beta(g) \le  \half \int _S H d A \,. \label{Paiva}  
\een
Combining \'Alvarez Paiva's (\ref{Paiva})  with (\ref{shell}) establishes
the conjecture (\ref{conjecture}) in this case.

In fact the proof is close to the ideas in \cite{Tod} and
so we briefly review it. If ${\bf n}$ is a unit vector
we define the height function on $S \subset {\Bbb E} ^3$ by
\ben
h= {\bf n}.{\bf x} \,, \qquad {\bf x \in S}\,.  
\een
Let $S_{\bf n}$ be the orthogonal
projection of the body $S$ onto a plane 
 with unit normal ${\bf n}$ and let $C( {\bf n}) =l( \partial
 S_{\bf n} ) $ be the 
perimeter of  $S_{\bf n}$. 
Then
\ben
\beta (g) \le \beta(h) \le C({\bf n}) \,. \label{height}  
\een

Now \cite{Tod}
\ben
\int _S H d A = { 1 \over 2 \pi} \int _{S^2} C({\bf n}) d \omega \,,
\label{cross} 
\een 
where $d \omega$ is the standard volume element on the round two-sphere 
$S^2$ of unit radius. Thus averaging (\ref{height}) over $S^2$ and using 
(\ref{cross}) gives (\ref{Paiva}).

\section{Shadows and widths}

The total mean curvature of a convex surface in Euclidean space has a 
number of interpretations. The width $w({\bf n}) = w(-{\bf n})$
is the distance between two parallel tangent planes with normals
$\pm {\bf n}$. One has
\ben
 {1  \over 2 \pi} \int_S  H dA  =  \langle w  \rangle = 
{ 1 \over 4 \pi} \int _S^2 w({\bf n} ) d \omega\,.  
\een    
Thus if
\ben
M={1 \over 8 \pi} \int _S H dA \,, 
\een
then 
\ben
W \ge 4 M \ge w  
\een
where $W$ is the greatest and $w$ the smallest width

Similarly the Tod points out \cite{Tod} that
\ben
{ 1  \over 16 \pi} C_m  \le M \le  {1 \over 4 \pi } C_m \,, 
\een
where $C_m$ is the largest perimeter of any orthogonal projection
of the body. 

\section{Quasi-local masses}

Recent suggestions for a quasi-local mass expression
 \cite{Liu1,Liu2,Wang1,Wang2,Wang3} have involved isometrically embedding
the horizon into Euclidean space ${\Bbb E}^3$. This will, by 
results of  Weyl and Pogorelov,
certainly be possible if the Gauss curvature of $S$ is positive.
In that case, since  the embedding is isometric, the Birkhoff invariant
can be calculated as if the surface is in flat Euclidean space and
the inequality  (\ref{Paiva}) holds. The total mean curvature
$H$ associated with the embedding into ${\Bbb E}^3$ also 
enters into the suggested expression for the quasi-local mass
of a trapped surface,
\ben
4 \pi M_{KLY} = \int _ S \bigl (H- \sqrt{2 \rho \mu} \bigr )   dA 
\een
where $-2\rho$ and $2\mu$ are the expansions of
outward and ingoing null normals, suitably normalised. 
  
Thus it is possible that some progress could be made there.
However, as we have seen above the Gaussian curvature of the Kerr-Newman
horizon can become negative near the poles, and as a consequence  it cannot
be isometrically  embedded into  ${\Bbb E}^3$.
Ignoring this difficulty for the time being,
we observe that in the time symmetric case for a marginally 
outer trapped surface
 
\ben
4 \pi M_{KLY} = \int _ S  H  dA  \le \beta (g) \,.
\een

\section{Areas} 

It is now well established that the area $A(g) $ of the outermost 
marginally trapped surface should satisfy Penrose's isoperimetric type 
conjecture that 
\ben
\sqrt {  \pi A(g)  } \le 4 \pi M_{ADM}\,. \label{cosmic} 
\een 

Evidently, if we could bound $\beta(g)$ 
above by  by $\sqrt {  \pi A(g)  }  $ 
we would have a proof of my version (\ref{conjecture}) of the  hoop conjecture.
On the other hand, if we can bound $\sqrt {  \pi A(g)  }  $
above by $\beta(g)$, then the  hoop conjecture  
would imply the Penrose conjecture. 

This raises the question of what 
is known about bounds for $A(g)$, $\beta(g)$, $l(g)$  and other 
invariants, either for a surface in general, or 
one with some additional restrictions.

We begin by noting that the Riemannian metric $g$ on $S$
allows us to define a distance $d(x,y)=d(y,x) , x, y \in S$ which is the 
infinum
of the length of all curves from $x$ to $y$. Then
\ben
b(x) = \max _y \,d(x,y) 
\een 
is the furthest  we can get from $x$. We then define
\bea
e(g) &=&\min _x b(x) = \min_x \max _y  d(x,y) \\
E(g) &=& \max _x b(x) = \max_x \max _y \, d(x,y)   
\eea

Hebda \cite{Hebda}  provides a lower bound for $A$: 
\ben
\sqrt{A(g)}\ge {1 \over \sqrt 2  }  \bigl(  2e(s) - E(g) \bigr )\label{hebda} \,.
\een
Using (\ref{cosmic}) we get
\ben
4 \pi M_{ADM }\ge {\sqrt {\pi \over 2}  } \bigl(  2e(s) - E(g) \bigr ) \,, 
\een
For the sphere the right hand side of (\ref{hebda}) is  $\sqrt{2 \pi ^3 } M_{ADM}$ which is satisfied but not sharp. There seems therefore no reason
to choose $C(g)= {\sqrt {\pi \over 2}  } \bigl(  2e(s) - E(g) \bigr )$,
in order to sharpen  Thorne's conjecture.

Another lower bound for the area has been given by Croke \cite{Croke}.
If, as above,  $l(g)$ is  the length of the shortest non-trivial geodesic
on $S$, then Croke proves that  
\ben
\sqrt{A(g)}\ge {1 \over 31} l(g) \,.
\een
This is again, far from the best possible result, which Croke conjectures to be
\ben
\sqrt{A(g)}\ge {1 \over 3^{1 \over 4} 2 ^{\half}} l(g)\,, \label{Croke}
\een
which is attained for two flat equilateral triangles glued back to back. 

If we use (\ref{cosmic} ) and (\ref{Croke}  we obtain
\ben
\Bigl (  { \pi ^2 \over 12} \Bigr )^{1 \over 4} l(g) \le 4 \pi M_{ADM} \label{croke2} \,.
\een

If one takes $C(g)=l(g)$, then  (\ref{croke2}) is 
weaker than Thorne's suggestion  and taking $C(g)= 
\Bigl ( { \pi ^2 \over 12} \Bigr )^{1 \over 4} l(g) $ 
looks rather perverse, and in any case there is a problem
about when it is attained. Moreover, since $\beta(g) \ge l(g)$, we cannot
easily relate (\ref{croke2}) to my form of the conjecture (\ref{conjecture}). 
Curiously however, for a special class of surfaces,
we can improve considerably on (\ref{hebda}) or  (\ref{croke2}).

\subsection{Horizons admitting an anti-podal map}

Many  results for general surfaces    rely on 
on the existence of  non-null homotopic closed curves.
For a surface with spherical topology 
no such curves exist. However it is possible to restrict
attention to the special class of surfaces  for which ${\Bbb Z}_2$ acts
freely and isometrically such that $x \rightarrow Ix$. 
The quotient $S^2/I \equiv {\Bbb R P} ^2$ and Pu provides 
a lower bound for$A(S/I)$ in terms of the 
the systole  ${\rm sys}(S/I )$, i.e. the length
of the shortest non-null homotopic curve:
\ben
\sqrt{A (S/I)  } \ge \sqrt{ 2 \over \pi} {\rm sys}(S/I )\,.
\een  
       
Now the shortest non-null homotopic curve on $S/I$
is a closed geodesic which lifts to a closed geodesic 
of twice the length 
on $S$, thus 
\ben 
{\rm sys}(S/I ) = \min _x \,d(x,Ix)  \le  b(x) \le e(g)\,,
\label{lower}  
\een
where $b(x)$ and $e(g)$ are taken on the spherical double cover.
If, as before,  $l(g)$ is  the length of the shortest non-trivial geodesic
on $S$, then for this class of metrics 
\ben
\sqrt{A(g)} \ge {2 \over \sqrt \pi} \,  {\rm sys}(S/I ) \ge { l(g) \over \sqrt \pi}
\een
and hence, using  (\ref{cosmic}) we obtain for this class of metrics,
\ben
l(g) \le 4 \pi M_{ADM}\,, 
\een  
i.e. the inequality (\ref{consequence}) which is a {\it consequence}
of my version of the hoop conjecture (\ref{conjecture}). 
Thus no counter example to to my conjecture can be constructed within
the class of horizons admitting an antipodal isometry.   

Of course (\ref{consequence})  is of the form of Thorne's suggestion, 
if we take  the 
circumference $C=l(g)$. However $l(g)$ does not 
carry with it the idea of the least circumference in all directions.
I have argued above  that it  is $\beta(g)$ which better captures that notion,
and so I prefer to think of (\ref{consequence}) of the more basic
inequality (\ref{conjecture}) and the fact that  (\ref{consequence})
holds in this special case as a confirmation of the general
plausibility of this line of argument.

It will perhaps be felt instructive  to recall some of the 
details of Pu's proof. He
makes use of the fact that any metric $g$  on ${\Bbb R} {\Bbb P}
^2$ may be written as
\ben
ds^2 = \Omega ^2(\theta, \phi) \bigl ( d \theta ^2 + \sin^2 \theta
d \phi ^2  \bigr ),
\een
where $\Omega(\theta, \phi)= \Omega(\pi-\theta, \phi + \pi)$. Thus
\ben
A(S/I) )  = \int \Omega ^2 \sin \theta d \theta d \phi ,
\een
and
$$
{\rm sys}(g) = 2 \inf_ x \inf _\gamma  \int _\gamma \Omega d \sigma
$$
where $\gamma$ is a curve running from $x \equiv (\theta, \phi) $
to $-x \equiv (\pi-\theta, \phi +\pi)$ and $d \sigma$ is the element of length
calculated using the round metric.

Now Pu considers  the effect of averaging the conformal factor
$\Omega$ with respect to the action of $SO(3)$, using the Haar or
bi-invariant measure on $SO(3)$. If the averaged metric, which is
of course the round metric,  is $\overline g$, one has
\ben
A({\overline g}) \le A(g)
\een
but
\ben
s({\overline g}) \ge s(g).
\een
Thus  the ratio $A/s^2$ is never smaller than for the round metric
${\overline g}$, and this case it is ${1\over \pi}$ and so his
inequality follows.

\subsection{Injectivity and Convexity Radii}

The mathematical literature on area, the lengths of geodesics etc
is often  couched in terms of the injectivity radius $i(g)$ and 
the convexity radius $c(g)$. In the sequel we mainly follow the
papers of Berger \cite{Berger1,Berger2,Berger3}. 
The definitions are valid for any dimension.    

The injectivity radius $i(x)$ of a point $x\in S$ is the supremum
of the distances out to which  which the exponential map is a
diffeomorphism onto its image. The injectivity radius $i(g)$ of the
manifold is the infinum over all points in $S$ of $i(x)$. In the
case of an axisymmetric body for which the metric may be written as
\ben
ds ^2 = R^2 \Bigl \{ d \theta ^2 + a^2(\theta) d \phi ^2 \Bigr \}  
\een
with $R$ an overall constant setting the scale, $0\le\theta \le \pi$,  
the injectivity radius of the north ($\theta=0$)  or
south ($ \theta =\pi $pole is $
{1 \over 2} l_p = \pi R$     and
\ben
i(g)  \le { 1 \over 2} l_p \,.
\een
Now local extrema of $a(\theta)$ correspond to azimuthal geodesics.
If the Gaussian   curvature is positive, there will only be one,
and   define $l_e$ as its length. Otherwise $l_e$ as the smallest
such length. 
\ben
l(g) \le l_p\,,\qquad l(g) \le l_e  
\een

The convexity radius $c(x)$ of a point $x$ is the largest radius
for which the geodesics ball $ B_c(x)$ centred on $x$ is geodesically convex,
that is every point in $B_c(x)$ is connected by a unique geodesic interval
lying entirely within $B(x)$. The
convexity radius $c(g) $ of the manifold is the infinum over all
points in $S$ of $c(x)$. On the round
unit-sphere ($R=1$) we have $i=\pi$ and $c ={\pi \over 2}$.

Now Berger proves that
\ben
l(g) \ge 2 c(g) \,.
\een
and hence 
\ben
\beta(g) \ge 2 c(g) \label{bong}
\een
Thus in the case of horizons admitting an  antipodal 
map, we can combine  Pu's result  
and (\ref{cosmic})  
to obtain
\ben
2c(g) \le 4 \pi M_{ADM} \,,
\een
in the case that my form of the hoop conjecture  (\ref{conjecture})
holds we obtain from (\ref{bong})  the same result.
In fact Klingenberg has shown that either
\ben
l(g) = 2 i(g) 
\een
or there is a geodesic segment of length $l(g)$ 
whose end points are conjugate.  
Finally for metrics on $S^2$ we have \cite{Berger4,Croke2}
the so-called {\it isembolic inequality}
\ben
\sqrt {\pi  A}  \ge 2 i(g) \ge 4 c(g)     
\een
and hence by (\ref{cosmic}) 
\ben
4 \pi M_{ADM}  \ge 2 i(g) \ge 4 c(g) \,.
\een

\section{Higher dimensions}

Birkhoff's invariant has a natural generalisation
to higher dimensions (see e.g. \cite{Colding} )
In the simplest case of an $n$-dimensional horizon, topologically
equivalent to $S^n$, one considers  foliations
whose leaves are topologically $S^{n-1}$'s.
There is then an obvious generalisation of (\ref{conjecture})  
relating the ADM mass  to the infinum over all foliations 
of the $(n-1)$-volume of the leaf of greatest $(n-1)$-volume.   

Pu \cite{Pu}  points out the obvious   generalisation to metrics on ${\Bbb R
} {\Bbb P} ^n$ which are conformal to the round metric. However
except in the case $n=2$, not every metric on ${\Bbb R} {\Bbb P}
^n$ is conformal to the flat round metric and so for $n>2$ this is
a very special case. However Berger's isembolic
inequality does generalise in an obvious way to all dimensions.

Of course  it is by now notorious \cite{Emparan} that
in five dimensional spacetimes, horizons need not be topologically
spherical  and in particular  one has black rings with horizon
topology  $S^1 \times S^2$. The methods and ideas of \cite{Colding}
should also be relevant in that case.     

Further discussion of the higher dimensional situation
will be deferred for  a future publication.

\section{Acknowledgements}  I am grateful to Gabor Domokos
for bringing \cite{Paiva} to my attention and hence re-igniting
my interest in these questions. I thank Harvey Reall, 
Gabriel Paternain, Chris Pope, Paul Tod
and Shing-Tung Yau and Claude Warnick for helpful comments and suggestions.

\end{document}